\documentclass[twocolumn,conference,10pt]{IEEEtran}
\usepackage{epsfig,amsfonts,subfigure,psfrag}
\usepackage{graphicx,cite,amssymb,amsmath}
\usepackage[nolist]{acronym}
\IEEEoverridecommandlockouts

\begin{document}

\title{High-Throughput Random Access via Codes on Graphs}

\author{
\authorblockN{Gianluigi Liva}
\authorblockA{Institute of Communication and Navigation\\
Deutsches Zentrum fÄur Luft- und Raumfahrt (DLR)\\
82234 Wessling, Germany\\
Gianluigi.Liva@dlr.de}
\and
\authorblockN{Enrico Paolini and Marco Chiani}
\authorblockA{DEIS, WiLAB\\
University of Bologna\\
via Venezia 52, 47023 Cesena (FC), Italy\\
\{e.paolini,marco.chiani\}@unibo.it}}
\date{\today}
\maketitle
\date{\today}
\thispagestyle{empty}

\setcounter{page}{0}

\maketitle

\begin{abstract}
Recently, contention resolution diversity slotted ALOHA (CRDSA) has
been introduced as a simple but effective improvement to slotted
ALOHA. It relies on MAC burst repetitions and on interference cancellation to increase the
normalized throughput of a classic slotted ALOHA access scheme.
CRDSA allows achieving a larger throughput than slotted ALOHA, at
the price of an increased average transmitted power. A way to
trade-off the increment of the average transmitted power and the
improvement of the throughput is presented in this paper. Specifically, it is proposed to divide
each MAC burst in $k$
sub-bursts, and to encode them via a $(n,k)$ erasure correcting
code. The $n$ encoded sub-bursts are transmitted over the MAC
channel, according to specific time/frequency-hopping patterns.
Whenever $n-e\geq k$ sub-bursts (of the same burst) are received
without collisions, erasure decoding allows recovering the remaining
$e$ sub-bursts (which were lost due to collisions). An interference
cancellation process can then take place, removing in $e$ slots the
interference caused by the $e$ recovered sub-bursts, possibly
allowing the correct decoding of sub-bursts related to other bursts.
The process is thus iterated as for the CRDSA case.
\end{abstract}


\section{Introduction}\label{sec:Intro}
Although the adoption of demand assignment multiple access (DAMA)
medium access control (MAC) protocols guarantee an efficient usage
of the available bandwidth, random access schemes remain an
appealing solution for wireless networks. Among them, slotted ALOHA
(SA) \cite{Abramson:ALOHA,Roberts72:ALOHA} is currently adopted as
initial access scheme in both cellular terrestrial and satellite
communication networks. In \cite{Rappaport83:DSA} an improvement to
SA was proposed, which is named diversity slotted Aloha (DSA). DSA
introduces a burst repetition which, at low normalized loads,
provides a slight throughput enhancement respect to SA. A more
efficient use of the burst repetition is provided by contention
resolution diversity slotted Aloha (CRDSA)
\cite{DeGaudenzi07:CRDSA}.

The idea behind CRDSA is the adoption of interference cancellation
(IC) for resolving collisions. More specifically, with respect to
DSA, the twin replicas of a burst (transmitted within a MAC
frame)\footnote{According to \cite{DeGaudenzi07:CRDSA}, in this
paper we consider a random access scheme where the slots are grouped
in MAC frames. We further restrict to the case where each user
proceeds with only one transmission attempt (either related to a new
packet or to a retransmission) within a MAC frame.} possess a
pointer to the slot position where the respective copy was sent.
Whenever a clean burst is detected and successfully decoded, the
pointer is extracted and the interference contribution caused by the
burst copy on the corresponding slot is removed. This procedure is
iterated, possibly permitting the recovery of the whole set of
bursts transmitted within the same MAC frame. This results in a
remarkably improved normalized throughput $T$,\footnote{$T$ is
defined as probability of successful packet transmission per time
slot.} which may reach $T\simeq0.55$, whereas the peak throughput
for pure SA is $T=1/e\simeq 0.37$. Further improvements may be
achieved by exploiting the capture effect
\cite{Roberts72:ALOHA,DeGaudenzi09:CRDSA}.

In \cite{Liva_CRDSA10_SCC} irregular repetition slotted Aloha was
introduced as an improvement to CRDSA, which permits to achieve
$T\simeq 0.8$. IRSA allows a variable repetition rate for each
burst. It is furthermore proved that, under the assumption of ideal
channel estimation and a sufficiently large signal-to-noise ratio
(SNR), the iterative burst recovery process can be represented via a
bipartite graph. It turns out that such a representation shares
several commonalities with the graph representation of the erasure
recovery process of low-density parity-check (LDPC)
codes\cite{Luby98:AndOr,studio3:Richardson_design_of_cap_appr}.
Since CRDSA is a specific instance of the IRSA approach, we will
refer in general to IRSA.

The performance improvement achieved by CRDSA/IRSA has nevertheless
a counterpart in the increment of the average transmitted power.
Focusing on CRDSA, each burst is replicated on the channels two
times. By assuming the same peak transmission power of the SA
scheme, the average power used for the transmission is doubled.

In this paper, we introduce a further generalization of IRSA (and
hence of CRDSA). The generalization is named coded slotted Aloha
(CSA) and works as follows. Each burst (of duration $T_{SA}$) is
divided in $k$ sub-bursts. The $k$ sub-bursts are then encoded by a
linear $(n,k)$ packet-level erasure correcting code. Throughout the
paper we rely on the assumption that the code is maximum distance
separable (MDS), if necessary constructed on a non-binary finite
field. Each of the so-obtained $n$ sub-bursts has a duration
$T_{CSA}=T_{SA}/k$. Keeping the overall MAC frame duration to
$T_{F}$, and neglecting the guard times, the MAC will be composed by
$N_{CSA}=kN_{SA}$ slots. The $n$ coded sub-bursts are thus
transmitted over $n$ slots picked at random. At the receiver side,
sub-bursts which collided with those sent by another user are marked
as lost. However, under the MDS code hypothesis, if the overall
number of lost sub-bursts of a specific burst are less than or equal
to $n-k$, the complete set of $n$ sub-bursts can be recovered.
Assuming that each sub-burst contains a pointer to the position of
the other sub-bursts related to the same burst, it is possible to
apply the IC process as for IRSA. Note that, apart from the
application of IC, a similar approach was already proposed in
\cite{Lam1990:THFHMA}, which will be referred to as time-hopping
multiple access (THMA).

Throughout the paper, we will illustrate how CRDSA and IRSA actually
represent particular cases of CSA. We will also explain how the
bipartite graph description still holds, but can be no longer
related to the simple bipartite graph of an LDPC code. A more
general type of bipartite graph description is required, which is
actually equivalent to that of double-generalized (DG) LDPC codes
\cite{paolini10:DGLDPC_random}.

\begin{figure}[]
\centerline{\includegraphics[width=\columnwidth]{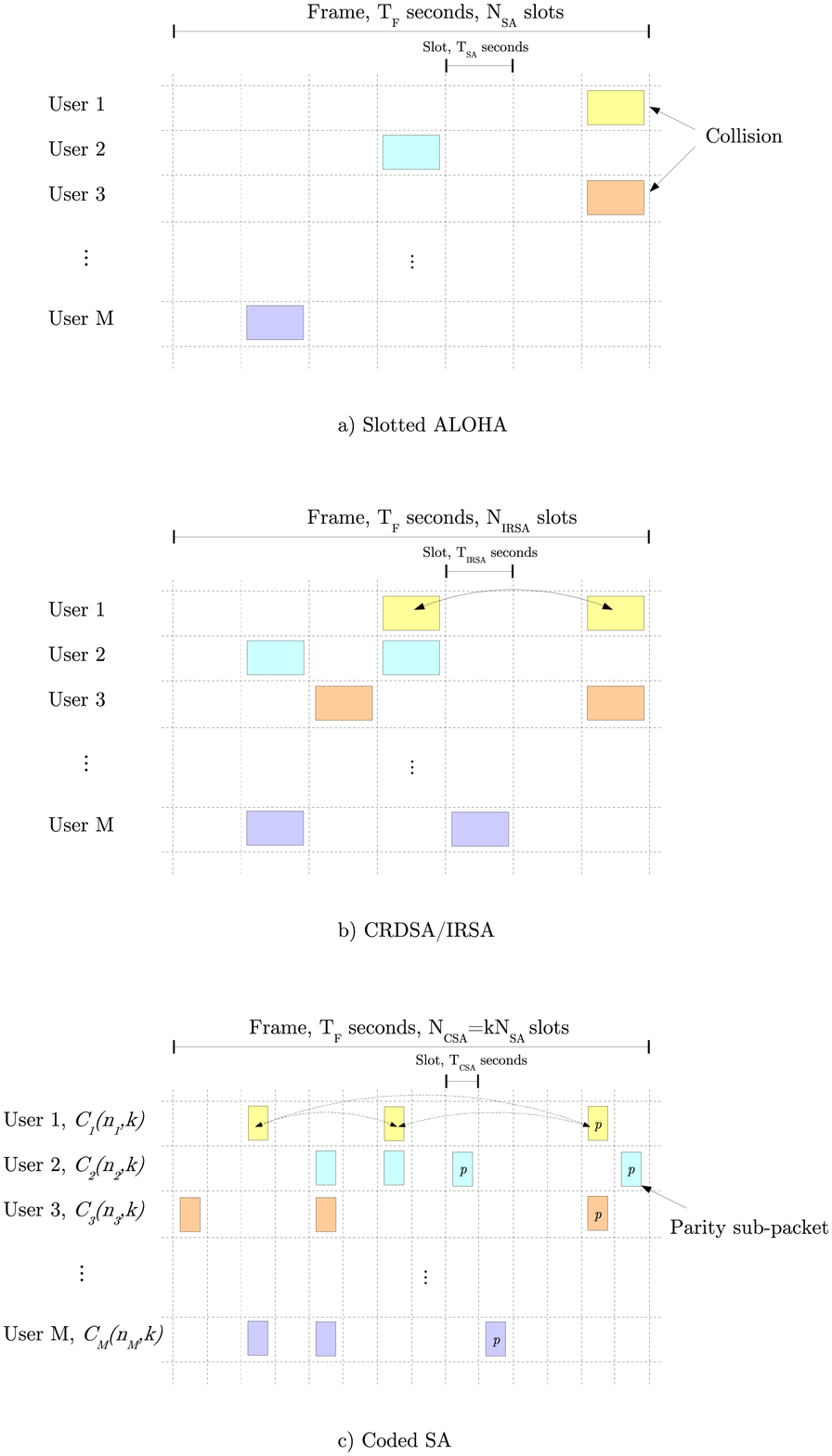}}
\caption{Slotted ALOHA (a), CRDSA/IRSA (b) and CSA (c) protocols
(framed). }\label{fig:system_overview}
\end{figure}

\begin{figure}[]
\centerline{\includegraphics[width=0.5\columnwidth]{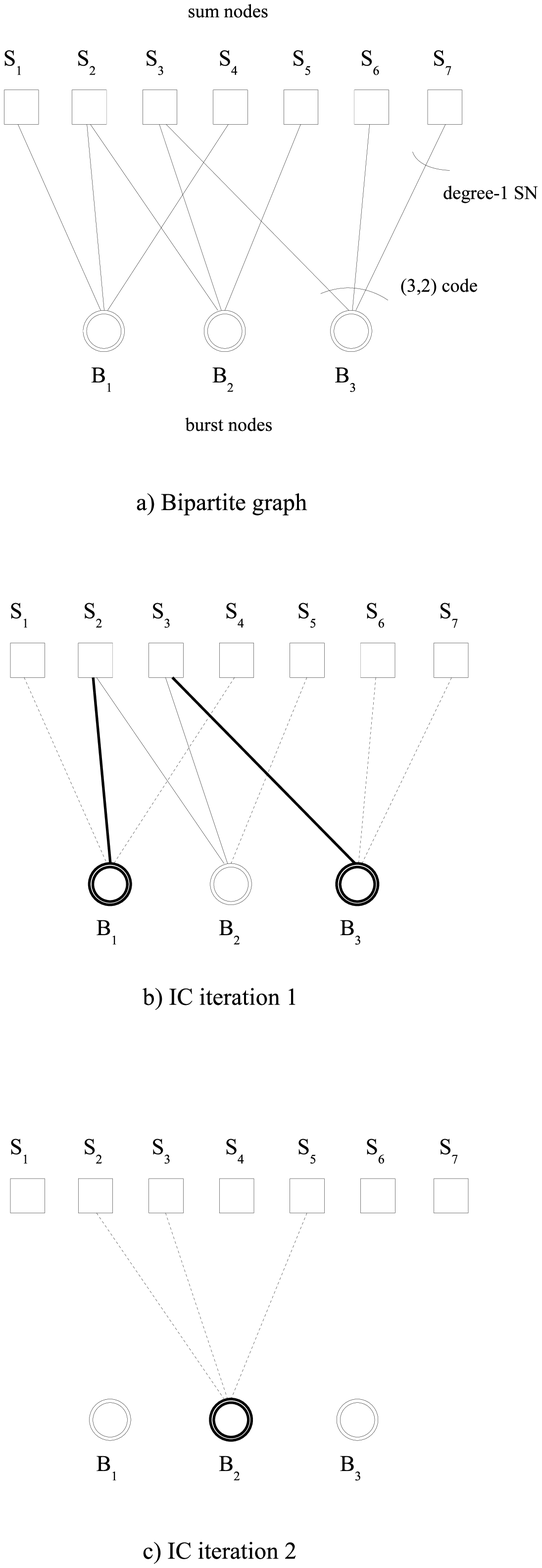}}
\caption{Graph representation of the IC iterative
process.}\label{fig:example}
\end{figure}

\section{System Overview}\label{sec:sys_over}
We will consider MAC frames of duration $T_F$. When SA or IRSA are
used, each MAC frame is composed of $N_{SA}=N_{IRSA}$ slots of
duration $T_{SA}=T_{IRSA}=T_F/N_{SA}$. The transmission of a packet
(or burst)\footnote{The notation \emph{burst} and \emph{packet} will
be interchangeably used to denote layer-2 data units.} is enforced
within one slot. We will assume that in each MAC frame a finite
number ($M$) of  users attempt a packet transmission. Without loss
of generality, each of the $M$ users performs a single transmission
attempt within the MAC frame, either related to a new packet or to a
retransmission of a previously collided one. Furthermore,
retransmissions shall not take place within the same MAC frame where
the collision happened. Hence, among the $M$ users, some may be
back-logged.  The normalized offered traffic (or channel traffic)
$G$ is given by $G=M/N_{SA}$, and represents the average number of
packet transmissions per SA slot. Pictorial representations of the
SA and of the IRSA techniques are depicted in Fig.
\ref{fig:system_overview}(a) and  in Fig.
\ref{fig:system_overview}(b). Considering IRSA (Fig.
\ref{fig:system_overview}(b)), note that each burst is replicated
within the MAC frame. Within each burst, pointers to its replicas
are provided so that, once a burst is recovered in a collision-free
slot, it is possible to extract the pointers to its replicas. If the
replicas collided in their slots, one could subtract from the
received signal their contribution, possibly allowing the recovery
of other bursts. In the example of Fig.
\ref{fig:system_overview}(b), a replica of the burst transmitted by
user 3 is received in the third slot. The contribution of the other
replica can be thus removed from the $7$th slot, permitting to
recover the burst sent by the $1$st user. The procedure can be
iterated, recovering the bursts of the other users.

When CSA is used, each burst is divided in $k$ sub-packets (called
\emph{units} in the following). The $k$ units are thus encoded by a
$(n,k)$ packet-level linear block code, resulting in $n>k$ units.
The MAC frame is consequently organized in $N_{CSA}=kN_{SA}$ slots,
each of duration $T_{CSA}=T_{SA}/k$. The $n$ units related to each
burst are then sent in $n$ different slots (within a MAC frame)
according to a time-hopping pattern. At the receiver side, units
which collided are declared as lost. Considering a specific user, if
$e$ units have been lost and $e\leq n-k$, the packet-level decoder
recover them. This approach, which is discussed in
\cite{Lam1990:THFHMA}, allows achieving a throughput higher than
that of SA when the offered traffic is moderate-low. In this paper,
we show how further improvements can be obtained by performing an IC
as it is done in IRSA. Note that, in principle, each user can make
use of a different code. More specifically, the generic user $i$ can
adopt any $(n_i,k)$ linear block code, provided that all the users
bursts are fragmented into the same number ($k$) of units. In the
case of \ref{fig:system_overview}(c), $k=2$. The $1$st, the $3$rd
and the $M$th users encode the $2$ units by a $(3,2)$ code, while
the $2$nd one uses a $(4,2)$ code. Note that user $2$ experiences
$2$ collisions (in the $5$th and in the $7$th slots). Nevertheless,
assuming the $(4,2)$ code to be MDS, the lost units can be
recovered, and their contribution can be removed from the
corresponding slots. At this stage, the unit sent from user $3$ in
the $5$th slot can be decoded, allowing (together with the unit
received in the $1$ slot) reconstructing the third unit (in slot
$13$). The IC process can be iterated, and further units (thus
bursts) can be recovered. Although the adoption of a different code
for each user provides a degree of freedom that could be beneficial
for the system performance, we will here consider only the case
where the users encode their units with the same $(n,k)$ code. We
will further assume for our analysis that the code is MDS, i.e., the
lost units (related to the same burst) can be recovered whenever at
least $k$ of them are received. Note that the offered traffic is
still given by $G=M/N_{SA}=kM/N_{SA}$. It is easy to verify that
IRSA is a particular case of CSA with $k=1$, i.e., where the unit is
encoded by  $(n,1)$  repetition codes.

\section{Graph Representation of the IC Process}\label{sec:graphIC}
It is now convenient to introduce a graph representation of the IC
process. We keep on considering a MAC frame composed of
$N_{CSA}=kN_{SA}$ slots, in which $M$ users attempt a transmission.
The MAC frame status can be described by a bipartite graph,
$\mathcal{G}=(B,S,E)$, consisting of a set $B$ of $M$ \textit{burst
nodes} (one for each burst that is transmitted within the MAC
frame), a set $S$ of $N_{CSA}$ \textit{sum} \textit{nodes} (one for
each slot), and a set $E$ of edges. An edge connects a burst node
(BN) $b_i\in B$ to a sum nodes (SN) $s_j\in S$ if and only if a unit
among the $n$ of the $i$-th burst is transmitted in the $j$-th slot.
Loosely speaking, BNs correspond to bursts and SNs correspond to
slots. Similarly, each edge corresponds to a unit. Therefore, a
burst is represented by a BN with $n$ neighbors (i.e. a BN from
which $n$ edges emanate). A slot where $d$ replicas collide will
correspond to a SN with $d$ connections. The number of edges
connected to a node is referred as the \textit{node degree}.

As an example, the bipartite graph representing a MAC frame with
$N_{CSA}=7$ slots where $M=3$ transmission attempts take place is
depicted in Fig. \ref{fig:example}(a), where squares denote sum
nodes and circles burst nodes. Each burst is divided in $k=2$ units
and encoded by a $(3,2)$ single parity-check code.

In our analysis, we will rely on two assumptions. 1)
\textit{Sufficiently high signal-to-noise ratio}. This assumption
allows to claim that, whenever a burst is received in a clean slot,
the decoding error probability is negligible. 2) \textit{Ideal
channel estimation}. This assumption is required to perform ideal
IC, allowing (together with the first one) the recovery of collided
bursts with a probability that is essentially one.\footnote{Details
on the implementation of the IC mechanism and on the performance of
CRDSA with actual channel estimation can be found in
\cite{DeGaudenzi07:CRDSA}.} Under these assumptions, the IC process
can be represented through a message-passing along the edges of the
graph. Following the example presented above, in Fig.
\ref{fig:example}(b), the iterative IC process starts by decoding
the first and the third bursts (in both cases $2$ units out of $3$
are received). The units that have been recovered were colliding in
slots $2$ and $3$. Their contribution can be then removed, allowing
the recovery of the $2$nd burst (Fig. \ref{fig:example}(c)).

It is worth now to introduce some further notation, namely the
concept of \textit{node-perspective degree distribution}. The sum
node degree distribution is defined by $\left\{\Psi_d\right\}$,
where $\Psi_d$ is the probability that a sum node possesses $d$
connections. A polynomial representation of the node-perspective degree distribution is given by
$\Psi(x)=\sum_d \Psi_d x^d$. %
It will be shown that the SN degree distribution is fully defined by
the system load $G$ and by the $(n,k)$ code parameters. The average
number of collisions per burst is defined as $\sum_d d\Psi_d =
\Psi'(1)$. It is easy to verify that $ G=M/N_{SA}=\Psi'(1)k/n$.

Degree distributions can be defined also from an \textit{edge
perspective}. We define  $\rho_d$ as the probability that an edge is
connected to a sum node of degree $d$. It follows from the
definitions that
\[
\rho_d=\frac{\Psi_d d}{\sum_d \Psi_d d}.
\]
The polynomial representations of  $\left\{ \rho_d \right\}$ is given by $\rho(x)=\sum_d \rho_d
x^{d-1}$. The relation $\rho(x)=\Psi'(x)/\Psi'(1)$ directly follow from the
definitions above.

\section{Iterative IC Convergence Analysis}\label{sec:EXIT_IT}
Consider now a burst node encoded via a $(n,k)$ code. Denote by $q$
the probability that an edge is unknown, given that each of the
other $n-1$ edges has been revealed with probability $1-p$ during
the previous iteration step. The edge will be revealed whenever at
least $k$ of the other edges have been revealed. Hence,
\begin{equation}
q_i=\sum_{e=n-k}^{n-1}{n-1 \choose e}p_{i-1}^e
(1-p_{i-1})^{n-1-e}.\label{eq:qi_evol}
\end{equation}
where the subscript of $p,q$ denotes the iteration index. In a
similar manner, consider a sum node with degree $d$. According to
the notation introduced so far, $p$ denotes the probability that an
edge is unknown, given that all the other $d-1$ edges have been
revealed with probability $1-q$ in the previous iteration step. The
edge will be revealed whenever all the other edges have been
revealed. Hence, $1-p=(1-q)^{d-1}$ or equivalently
$p=1-(1-q)^{d-1}$. According to the tree analysis argument of
\cite{Luby98:AndOr}, by averaging the expression over the edge
distribution, one can derive the evolution of the average erasure
probabilities during the $i$-th iteration for the sum nodes as
\begin{equation}
p_{i}=\sum_d \rho_d
\left(1-\left(1-q_i\right)^{d-1}\right)=1-\rho\left(1-q_i\right).\label{eq:pi_evol}
\end{equation}
For sake of simplicity, the iteration index will be omitted in the
rest of the paper. By iterating these equations for a given amount
of times (limited to $I_{max}$), it is possible to analyze the
iterative convergence of the IC process. The initial condition has
to be set as $q_0=p_0=1$, i.e., there are no revealed edges at the
beginning of the process. According to (\ref{eq:pi_evol}), at the
first iteration $p$ will take the value given by the probability
that an edge is not connected to a degree-$1$ sum node. It is
important to remark that the recursions in (\ref{eq:qi_evol}) and
(\ref{eq:pi_evol}) hold if the messages exchanged along the edges of
the graph are statistically independent. Thus, the accuracy of
(\ref{eq:qi_evol}),(\ref{eq:pi_evol}) is subject to the absence of
loops in the bipartite graph (recall that loops introduce
correlation in the evolution of the erasure probabilities). This
assumption implies very large frame sizes ($M\rightarrow\infty$ and
consequently $N_{CSA}\rightarrow \infty$ for fixed $G$), and the
analysis presented next will refer to this asymptotic setting. This
hypothesis is nevertheless needed just for deriving a distribution
design criterion. It will be shown that distributions designed for
the asymptotic setting are effective also when considering realistic
MAC frame sizes.

By fixing the $(n,k)$ code, for each value of the offered traffic
$G$ the distribution $\rho(x)$ can be determined. For values of $G$
below a certain threshold $G^*$, the iterative IC will succeed with
probability close to $1$ (almost all the bursts will be recovered).
Above the threshold $G^*$, the procedure will fail with a
probability bounded away from $0$. We will look thus for codes
leading to an high threshold $G^*$, thus allowing (in the asymptotic
setting) error-free transmission for any offered traffic up to
$G^*$.

To complete the analysis, it is nevertheless required to know the
distribution $\rho(x)$ in (\ref{eq:pi_evol}). The probability that a
sum node is of degree $d$ is given by
\[
\Psi_d={M\choose d}
\left(\frac{\Psi'(1)}{M}\right)^d\left(1-\frac{\Psi'(1)}{M}\right)^{M-d}.
\]
The node-perspective sum nodes degree distribution results in
\begin{equation}
\begin{array}{ll}
  \Psi(x) & =\displaystyle\sum_d{M\choose d}
\left(\frac{\Psi'(1)}{M}\right)^d\left(1-\frac{\Psi'(1)}{M}\right)^{M-d}
x^d \\ &= \left(1-\frac{\Psi'(1)}{M}(1-x)\right)^M,
\end{array}
\label{eq:rhox_monster}
\end{equation}
By letting $M \rightarrow \infty$ (asymptotic setting),
(\ref{eq:rhox_monster}) becomes $$\Psi(x)=e^{-\Psi'(1)(1-x)}=e^{-G(1-x)n/k}\,.$$ The
edge-perspective sum nodes degree distribution is therefore given by
\begin{equation}
\rho(x)=\frac{\Psi'(x)}{\Psi'(1)}=e^{-G(1-x)n/k}.\label{eq:rho}
\end{equation}

\begin{figure}[]
\centerline{\includegraphics[width=\columnwidth,draft=false]{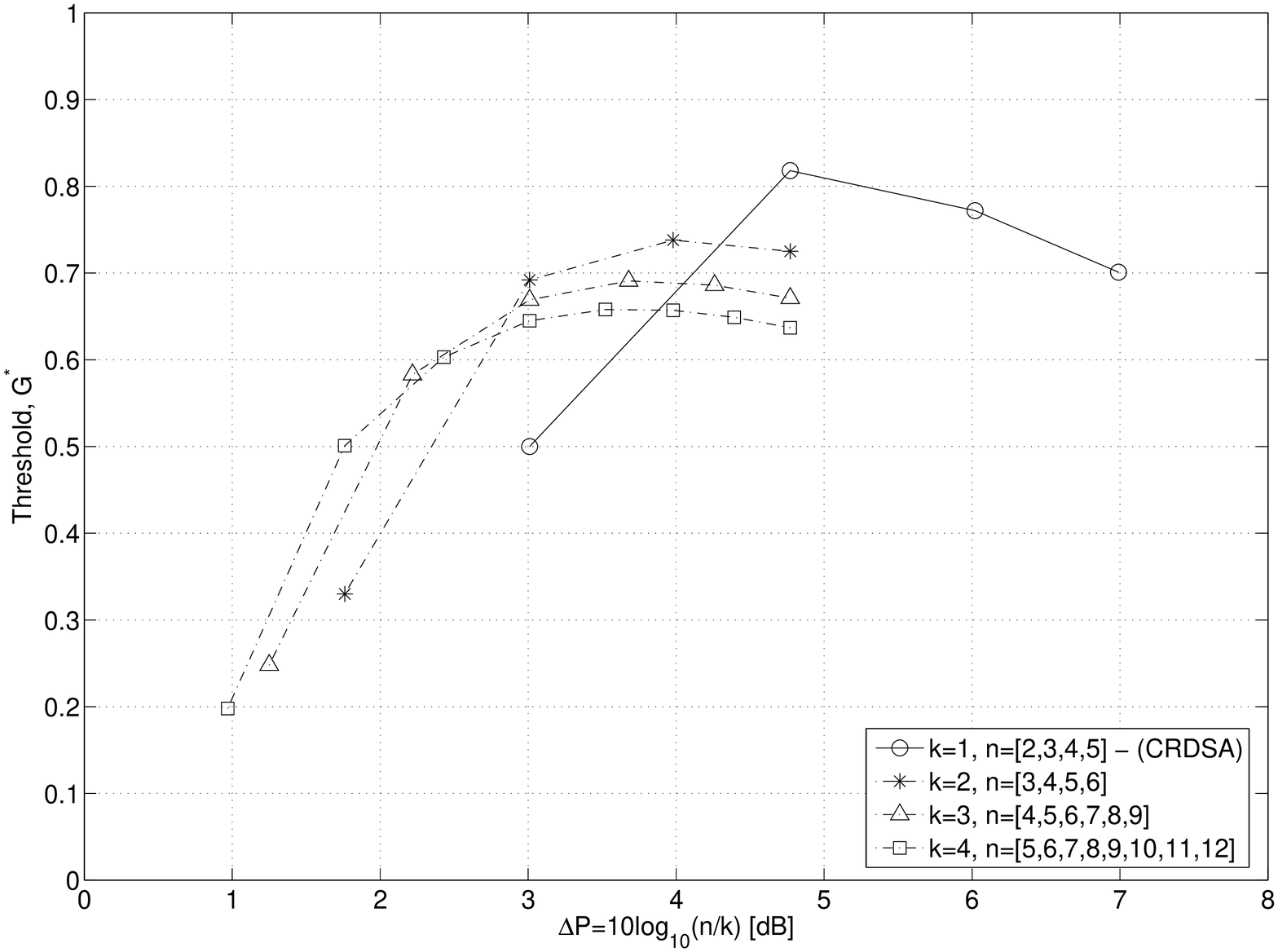}}
\caption{Comparison of the thresholds $G^*$ for different  $(n,k)$
codes, vs. the average power penalty $\Delta
P=10\log_{10}(n/k)$.}\label{fig:Comparison_Thresholds}
\end{figure}

Some thresholds for different $(n,k)$ codes are provided in Fig.
\ref{fig:Comparison_Thresholds}, as functions of the average power
increment w.r.t. SA (referred as average power penalty), which is
given by $\Delta P=10\log_{10}(n/k)$. Considering codes with rate
$k/n=1/2$, thus leading to a penalty of $3$ dB, we note that the
best threshold is obtained by CSA based on a $(4,2)$ code, for which
$G^*=0.692$. A repetition-$2$ CRDSA would provide a much lower
threshold ($G^*=0.5$). Note also that the same threshold $G^*=0.5$
can be obtained by using CSA with a $(6,4)$ code, thus saving more
than $1.2$ dB of average power.

A remark is deserved for the case where each user adopts a $(n,k)$
single parity-check code (i.e., $n=k+1$). In this case,
(\ref{eq:qi_evol}) can be simplified as $q=1-(1-p)^k$. Let's define $f(p)=1-(1-p)^k$ and let $g(p)$
be the inverse of
(\ref{eq:pi_evol}), i.e. $g(p)=-[k/(G(k+1))]\ln (1-p)$. One can then
derive a simple upper bound to $G^*$ by imposing ${\rm d}f(p)/{\rm d}p\leq{\rm d}q(p)/{\rm d}p$ for
$p\rightarrow 0$ and $G=G^*$. It follows that
\[
G^*\leq\frac{1}{k+1}.
\]
It is possible to check from Fig. \ref{fig:Comparison_Thresholds}
that in all the cases where $n=k+1$ such bound is tightly
approached.

\begin{figure}[]
\centerline{\includegraphics[width=\columnwidth,draft=false]{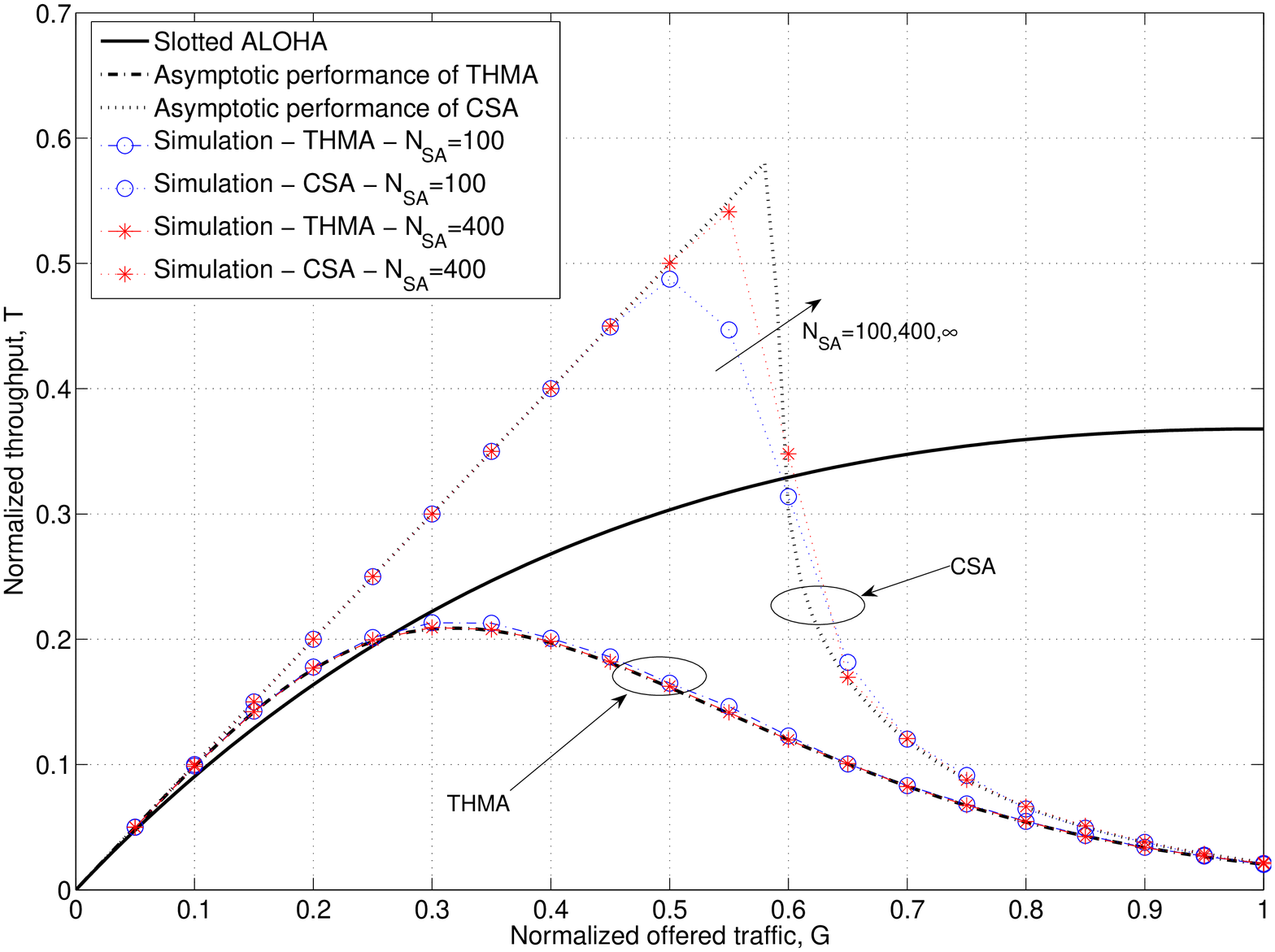}}
\caption{Simulated throughput for SA, THMA, and for CSA with $k=4$,
$n=7$ and $N_{SA}=100, 400, \infty$ ($I_{max}=20$).}
\label{fig:TG_Ns_Comparisons_k4n7_Imax20}
\end{figure}

\section{Numerical Results}\label{sec:Thrp_eval}
We focus next on the case where a $(7,4)$ code is used. For this
specific case, results of simulations and of IC analysis are
summarized in Fig. \ref{fig:TG_Ns_Comparisons_k4n7_Imax20}. In both
cases, we assumed a maximum amount of iterations $I_{max}=20$. The
simulation results are provided for $N_{SA}=100$ and $N_{SA}=400$,
while the analytical curves are relevant to the asymptotic setting
$N_{SA}\rightarrow \infty$. The performance of both THMA and of CSA
are provided. Some remarks follow.

\smallskip

\noindent 1. The match between the simulation results and the
proposed analytical approach is good. For sufficiently large MAC
frames, the simulated performance approaches the asymptotic curve,
providing an evidence of the validity of the proposed analysis.

\smallskip

\noindent 2. The gain due to the iteration of IC over THMA is
evident. While THMA achieves a peak throughput of about $0.2$, CSA
approaches $T\simeq 0.55$ for $N_{SA}=400$.

\smallskip

\noindent 3. When $G>G^*=0.6$, the performance of CSA drops quickly
below the SA performance, and converges for larger $G$ to the
performance of THMA. This is due to the fact that, when $G>G^*$, the
IC process gets stuck in an early stage, leaving most of the
collisions unresolved, and thus it does not improve much the
performance of THMA.

\smallskip

\noindent 4. The region where $T\simeq G$ goes up to $G\simeq 0.5$
for CSA with $N_{SA}=400$, meaning that up to such values the
offered traffic turns into useful throughput (i.e., the burst loss
probability is very small). For THMA, this holds just for $G<0.1$,
while for SA it hold for very small values of $G$.

\section{Conclusions}
In this paper, a generalization of the CRDSA/IRSA approach for MAC
has been introduced, namely, CSA. The generalization consists of
dividing each burst into sub-bursts and encoding them through a
linear packet erasure correcting code, which has been assumed to be
MDS. Iterative IC is then performed on the sub-bursts and combined
with the local erasure correction capability of each packet erasure
code. It has been highlighted how iterative IC in the context of CSA
is analogous to the iterative decoding over the erasure channel of
DG-LDPC codes based on sparse bipartite graphs. Numerical results
have been provided, confirming the effectiveness of the proposed
approach.

\section*{Acknowledgment}

This work was supported in part by the Europ. Comm. under project
FP7 ``OPTIMIX''.

\bibliographystyle{ieeetr}

\end{document}